# Environment-induced decay of teleportation fidelity of the one-qubit state

Ming-Liang Hu[*]

*School of Science, Xi'an University of Posts and Telecommunications, Xi'an 710061, China*

**Abstract:** The one-qubit teleportation protocol is reexamined when it is executed in the presence of various decohering environments. The results revealed that this quantum protocol is more robust under the influence of dephaisng environment than those under the influence of dissipative or noisy environment. The environment may deprive the quantum advantage of teleportation over purely classical communication in a finite or infinite lifetime, which is dependent on the type of environment. Also we found that except entanglement, the purity of the entangled state resource is also crucial in determine the quality of the teleported state.

As an archetype of quantum communication, quantum teleportation [1] has been studied by a number of authors both theoretically [2–5] and experimentally [6–9]. The key requirements for this quantum protocol are the performance of clean projective operations that process information and the prior shared maximally entangled state resource between the sender (traditionally named Alice) and the receiver (traditionally named Bob). This needs the processors as well as the transmission channels to be isolated perfectly from the surrounding environments. However, in reality, this may be a very difficult experimental task. For example, previous studies [10–13] have demonstrated that decoherence always induce degradation of entanglement, thus the preparation of the maximally entangled channel states is a hard work. Moreover, in large-scale realization of quantum communication based on optical systems, errors may occur due to the inevitable photon loss (an analog of an erasure in classical information theory) in the transmission channel [6], this limits the distance for efficient communication and is therefore to be recognized as a fundamental difficulty. Indeed, there are works [14–17] showing that the environmental effects may cause a teleportation to lose its quantum advantage over purely classical communication. Thus from a practical point of view, it is significant to understand the decoherence mechanism induced by various external environments and their influence on fidelity of quantum teleportation. This is also obviously vital for designing effective strategies [3,4,18] to implement high efficient and long distance quantum communication.

---

[*] Corresponding author.
 Tel.: +86 029 88166094
 *E-mail address*: mingliang0301@163.com, mingliang0301@xupt.edu.cn (M.-L. Hu)

The original protocol of teleportation proposed by Bennett et al. [1] is implemented through a channel involving an Einstein-Podolsky-Rosen (EPR) pair established previously between Alice and Bob (see Fig. 1). This protocol enables perfect teleportation of an arbitrary (pure or mixed) one-qubit state in the idealistic situation. However, in practical implementation of this protocol, there are several stages that the decoherence may take place and thus reduce the fidelity of the expected outcomes. First, the unknown state to be teleported may lose its coherence and becomes a mixed state before it is teleported; this case is trivial since Bob can always retrieve the same decohered state if the channel is protected perfectly from decohering environments [1]. Second, during the establishment of the shared channel state between Alice and Bob, decoherence can also take place. In real experiments, the entangled state resource used for teleportation can be prepared by a third party and then sends one qubit to Alice and the other one to Bob, or prepared by Alice who keeps one qubit with herself and sends another one to Bob. During the transmission process, the qubit(s) may be exposed to external environment, and degrade quantum entanglement between them. Moreover, while Alice performs the Bell basis measurements or Bob performs the recovery operations the decoherenece may also be set in [14,16].

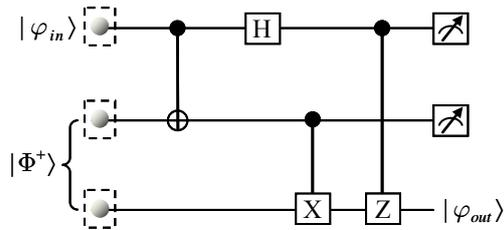

**Fig. 1.** A quantum gate circuit for teleportation in the presence of decohering environment. The top two qubits are in possession of Alice, while the bottom one is in possession of Bob. The "ammeter" symbols represent quantum measurements, and the dotted boxes denote decoherence imposed by external environments.

In this Letter we reexamine the one-qubit teleportation protocol [1] when it is executed in the presence of decohering environments as illustrated in Fig. 1. To describe the decoherence process mathematically, we assume that each individual particle of the system interacts independently with the environment, then under weak system-environment coupling and in the Markovian limit, the dynamics of the system can be described by the Lindblad master equation [19,20]

$$\frac{d\rho}{dt} = \sum_{k,i} \frac{\gamma_k}{2} (2\mathcal{L}_{k,i} \rho \mathcal{L}_{k,i}^\dagger - \{\mathcal{L}_{k,i}^\dagger \mathcal{L}_{k,i}, \rho\}), \qquad (1)$$

where the brace $\{\}$ means anticommutator. $\rho(t)$ and $\gamma_k$ denote, respectively, the reduced density operator of the system and the coupling strengths of the qubits with their respective environments. Different decoherence process may arise when a system is coupled to different environments, and they can be described in terms of the above master equation by a suitable choice of the generators of decoherence $\mathcal{L}_{k,i}$. In contrast to Refs. [14–16], here we consider $\mathcal{L}_{k,i}$ defined in terms of the raising and lowering operators $\sigma^\pm = (\sigma^x \pm i\sigma^y)/2$, where $\sigma^\alpha$ with $\alpha = x, y, z$ are the usual Pauli operators. We will discuss three different mechanisms of decoherence [10]. The first is the dissipative process with the dissipator given by $\mathcal{L}_k = \sigma_k^-$. This process describes coupling of the central system to a thermal bath at zero temperature. The second case is the infinite temperature environment described by generators of decoherence $\mathcal{L}_{k,1} = \sigma_k^-$ and $\mathcal{L}_{k,2} = \sigma_k^+$, in which decay and excitation occur at exactly the same rate, and the noise induced by the transitions between the two levels brings the system to a stationary and maximally mixed state. The third case we discuss is the purely dephasing reservoir with $\mathcal{L}_k = \sigma_k^+ \sigma_k^-$. Different from the dissipative and the noisy processes, this process introduces only loss of phase coherence and there are no changes in the populations of the ground and excited states.

To examine quantitatively the environmental effects on fidelity of quantum teleportation, it is convenient to express the unknown one-qubit state needs to be teleported as $|\varphi_{in}\rangle = a|0\rangle + b|1\rangle$, with $a = \cos(\theta/2)e^{i\phi/2}$ and $b = \sin(\theta/2)e^{-i\phi/2}$ ($0 \leqslant \theta \leqslant \pi$ and $0 \leqslant \phi \leqslant 2\pi$). When one adopts the maximally entangled Bell state $|\Phi^+\rangle = (|00\rangle + |11\rangle)/\sqrt{2}$ as quantum channel for teleportation in the presence of decoherence, the density operator for the output state is given by

$$\rho_{out} = \mathrm{Tr}_{1,2}\{\mathcal{U}_{tel}\varepsilon(\rho_{in} \otimes \rho_{en})\mathcal{U}_{tel}^\dagger\}, \qquad (2)$$

where $\rho_{in} = |\varphi_{in}\rangle\langle\varphi_{in}|$, $\rho_{en} = |\Phi^+\rangle\langle\Phi^+|$, and $\mathrm{Tr}_{1,2}$ is a partial trace over the two qubits belong to Alice (from hereon the three qubits shown in Fig. 1 will be designated by 1, 2 and 3 from top to bottom). $\varepsilon$ represents the quantum operation which maps the state of the system from $\rho_{in} \otimes \rho_{en}$ to $\varepsilon(\rho_{in} \otimes \rho_{en})$ due to its coupling with the decohering environment, and the explicit expressions for $\varepsilon(\rho_{in} \otimes \rho_{en})$ can be obtained by solving the appropriate master equation (1) with $\rho_{in} \otimes \rho_{en}$ as the initial state. Moreover, $\mathcal{U}_{tel} = \mathcal{C}_{13}^Z \mathcal{C}_{23}^X \mathcal{H}_1 \mathcal{C}_{12}^X$ is a unitary operator, where $\mathcal{H}_1$ stands for the Hadamard operation on qubit 1, and $\mathcal{C}_{mn}^{\alpha}$ ($\alpha = \text{X, Z}$) stands for the controlled-$\alpha$ operation with $m$ as the control qubit and $n$ the target qubit.

The quality of quantum teleportation in the presence of external environment can be evaluated by calculating the fidelity and average fidelity. The fidelity is defined as

$$F(\theta,\phi) = \langle \varphi_{in} | \rho_{out} | \varphi_{in} \rangle. \tag{3}$$

This quantity gives the information of how close the teleported state $\rho_{out}$ is to the unknown state $\rho_{in}$ to be teleported, i.e., they are equal when $F(\theta,\phi) = 1$ and orthogonal when $F(\theta,\phi) = 0$. Furthermore, by performing an average over all possible input states on the Bloch sphere one can get the average fidelity as

$$F_{av} = \frac{1}{4\pi} \int_0^{2\pi} d\phi \int_0^{\pi} d\theta \sin\theta F(\theta,\phi), \tag{4}$$

where $4\pi$ is the solid angle.

In the present work, we would like to investigate decay dynamics of average fidelity when the teleportation is executed in the presence of dissipative, noisy and dephasing environment [10]. For every type of environment, we will analyze the following three different cases. The first is that both the two qubits constitute the transmission channel subject to decoherence. This corresponds to the physical situation in which the entangled state resource used for teleportation is prepared by a third party, thus they must be exposed to decohering environments when being transmitted to Alice and Bob. The second case is that only the qubit of Bob subjects to decoherence. This may correspond to the situation that Alice prepares the entangled state resource and keeps one qubit with herself and sends another one to Bob. Alice can try her best to isolate her qubit perfectly from decoherence, but Bob's qubit has to face the environmental influence during the transmission process. Finally, we explore the case that both the two qubits at the transmitting station (i.e., the qubits of Alice) subject to decoherence, for which the quality of the teleported state is expected to be inferior to the former two cases because the unknown state to be teleported may also be destroyed and becomes a mixed state before it is teleported.

Together with different environmental types, one can obtain different average fidelities $F_{av}^{(\alpha,ca)}$. Here $\alpha = di$, $no$ or $de$ indicates the situation that the system subjects to dissipative, noisy or dephasing environment, while $ca = \{1, 2, 3\}$ designate, respectively, the case that the two qubits constitute the quantum channel, the qubit at Bob's possession or the qubits at Alice's possession subject to decoherence (same notations will be used throughout this Letter). Moreover, the system state at a given time, say $t$, can be obtained from Eq. (1) by choosing the system-environment coupling strengths as $\gamma_1 = 0$ and $\gamma_2 = \gamma_3 \equiv \gamma$ for $ca = 1$, $\gamma_1 = \gamma_2 = 0$ and $\gamma_3 = \gamma$ for $ca = 2$, $\gamma_1 = \gamma_2 = \gamma$ and $\gamma_3 = 0$ for $ca = 3$ (i.e., here we assume the same system-environment coupling strengths for the decohered qubits).

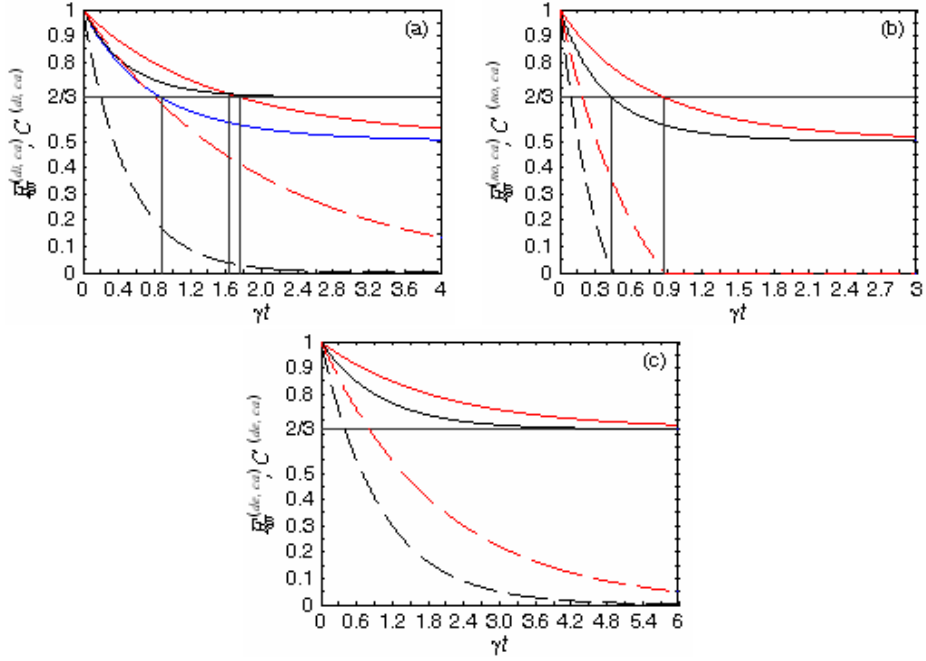

**Fig. 2.** (Color online) Average fidelity of teleportation (solid curves) and concurrence of the channel state (dashed curves) versus $\gamma t$ for the case of the system subjects to dissipative (a), noisy (b) and dephaisng (c) environments. For every line type, the black, red and blue curves correspond to the case of $ca = 1$, $2$ and $3$. Note that here the curves for $C^{(\alpha,2)}$ and $C^{(\alpha,3)}$ ($\alpha = di, no, de$) as well as for $F_{av}^{(\alpha,1)}$ and $F_{av}^{(\alpha,3)}$ ($\alpha = no, de$) are overlapped.

First, we discuss the situation in which the system subjects to dissipative environment. The system state $\varepsilon(\rho_{in} \otimes \rho_{en})$ composed of the state to be teleported and the quantum channel can be derived by solving the master equation (1) with the generators of decoherence given by $\mathcal{L}_k = \sigma_k^-$. After a straightforward algebra, we obtain

$$F_{av}^{(di,1)} = \frac{2}{3} + \frac{1}{3}e^{-2\gamma t},$$
$$F_{av}^{(di,2)} = \frac{1}{2} + \frac{1}{3}e^{-\frac{1}{2}\gamma t} + \frac{1}{6}e^{-\gamma t}, \qquad (5)$$
$$F_{av}^{(di,3)} = \frac{1}{2} + \frac{1}{3}e^{-\gamma t} + \frac{1}{6}e^{-2\gamma t}.$$

From the average fidelities expressed in Eq. (5), one can see that for the case that both the two qubits involved in the transmission channel subject to dissipative environment, the teleportation protocol always outperform those of purely classical ones since $F_{av}^{(di,1)}$ approaches to its limiting value $2/3$ [21] only when $\gamma t_c^{(di,1)} \to \infty$. For the cases that the qubit of Bob or the two qubits of Alice subject to dissipative environment, however, there is a finite lifetime $\gamma t_c^{(di,ca)}$ ($ca = 2, 3$) beyond which the teleportation protocol will fail to attain an average fidelity better than classically possible. From Eq. (5), one can derive the critical rescaled time $\gamma t_c^{(di,ca)}$ ($ca = 2, 3$) analytically as $\gamma t_c^{(di,2)} = 2\ln(\sqrt{2}+1)$ and $\gamma t_c^{(di,3)} = \ln(\sqrt{2}+1)$. Moreover, as can be seen from Fig. 2a, although the case with both the two qubits of the transmission channel subject to dissipative environment always yields the concurrence [22] $C^{(di,1)} = e^{-2\gamma t}$ smaller than that of $C^{(di,2)} = e^{-\gamma t/2}$ with only the qubit of Bob subjects to dissipative environment, the average fidelity $F_{av}^{(di,1)}$ becomes larger than $F_{av}^{(di,2)}$ when $\gamma t$ increases after $\gamma t > 2\ln[6/(m^{1/3} - 2m^{-1/3} - 2)] \simeq 1.6391$ ($m = 64 + 6\sqrt{114}$). This indicates that in general, channel state with greater amount (not maximum) of entanglement does not necessarily yield better teleportation fidelity. Physically, one could attribute the cause of this phenomenon to the fact that there is now a comparatively small purity [2] for the case that the qubit of Bob subjects to dissipative environment after a critical rescaled time. In fact, one can derive the purity of the channel state analytically as $P^{(di,1)} = 1 - 2e^{-\gamma t} + 3e^{-2\gamma t} - 2e^{-3\gamma t} + e^{-4\gamma t}$ and $P^{(di,2)} = 1/2 + e^{-2\gamma t}/2$, which gives rise to the following inequality $P^{(di,1)} > P^{(di,2)}$ in the region of $\gamma t > \ln[6/(n^{1/3} - 14n^{-1/3} + 2)] \simeq 0.9248$ ($n = 8 + 6\sqrt{78}$). Thus here we demonstrated that the concurrence may only describe certain aspects of the entanglement; it cannot completely reflect the quality of the entanglement as a resource for teleportation because the purity of the channel state is also very important.

For the situation in which the system subject to noisy environment, by repeating calculation of the preceding section with changing the generators of decoherence to $\mathcal{L}_{k,1} = \sigma_k^-$ and $\mathcal{L}_{k,2} = \sigma_k^+$ [10], we obtain the average fidelity as

$$F_{av}^{(no,1,3)} = \frac{1}{2} + \frac{1}{3}e^{-2\gamma t} + \frac{1}{6}e^{-4\gamma t},$$
$$F_{av}^{(no,2)} = \frac{1}{2} + \frac{1}{3}e^{-\gamma t} + \frac{1}{6}e^{-2\gamma t}.$$
(6)

Plots of the above equations are displayed in Fig. 2b as solid curves. Different from the former case, here all the average fidelities become smaller than $2/3$ after a rescaled critical time, which can be obtained analytically as $\gamma t_c^{(no,1,3)} = [\ln(\sqrt{2}+1)]/2$ and $\gamma t_c^{(no,2)} = \ln(\sqrt{2}+1)$. One can also obtain the concurrence of the channel state analytically as $C^{(no,1)} = \max\{e^{-2\gamma t} + e^{-4\gamma t}/2 - 1/2, 0\}$ and $C^{(no,2,3)} = \max\{e^{-\gamma t} + e^{-2\gamma t}/2 - 1/2, 0\}$. For the case that both the two qubits involved in the quantum channel or the qubit in possession of Bob subject to decoherence, the noisy environment also disentangling the channel state in a finite lifetime $\gamma t_c^{(no,1)}$ and $\gamma t_c^{(no,2)}$, respectively (see Fig. 2b), which is known as entanglement sudden death (ESD) observed previously by Yu and Eberly [23] and has been widely studied recently. Therefore, the success of the teleportation protocol is constrained by the finite lifetime of the entanglement of the channel state. For the case that the two qubits of Alice subject to noisy environment, the average fidelity has completely the same form as that with both the two qubits of the quantum channel subject to noisy environment, however, when $F_{av}^{(no,3)}(t_c) = 2/3$ we obtain $C^{(no,3)}(t_c) = (\sqrt{2}-1)^{1/2} + \sqrt{2}/2 - 1 \simeq 0.3507$. Thus for this special case, the quantum channel should possess a nonzero critical value of minimum entanglement in order to teleport the one-qubit state with fidelity better than the best possible score if Alice and Bob communicate with each other only via a classical channel. Moreover, we highlight that here the purity $P^{(di,1)} = 1/4 + e^{-4\gamma t}/2 + e^{-8\gamma t}/4$ is also smaller than $P^{(di,2,3)} = 1/4 + e^{-2\gamma t}/2 + e^{-4\gamma t}/4$ in the whole time region.

Now we turn our attention to the implementation of one-qubit teleportation with the system subject to dephaisng environment. By writing the generators of decoherence as $\mathcal{L}_k = \sigma_k^+ \sigma_k^-$ [10] and performing similar calculations as the preceding sections, we obtain the average fidelity as

$$F_{av}^{(de,1,3)} = \frac{2}{3} + \frac{1}{3}e^{-\gamma t},$$
$$F_{av}^{(de,2)} = \frac{2}{3} + \frac{1}{3}e^{-\frac{1}{2}\gamma t}. \quad (7)$$

From Eq. (7) one can see that when being executed in the presence of dephasing environment, the teleportation protocol always outperform those of purely classical ones since $F_{av}^{(de,ca)} > 2/3$ ($ca = 1,2,3$) in the whole time region [21]. Moreover, one can obtain explicitly the concurrence [22] and purity of the channel state as $C^{(de,1)} = e^{-\gamma t}$, $C^{(de,2,3)} = e^{-\gamma t/2}$, and $P^{(de,1)} = 1/2 + e^{-2\gamma t}/2$, $P^{(de,2,3)} = 1/2 + e^{-\gamma t}/2$, this, together with Eq. (7) indicates that the average fidelities always decrease with the decrease of entanglement and purity of the channel state, and in the long time limit we have $F_{av}^{(de,ca)} \to 2/3$, $C^{(de,ca)} \to 0$ and $P^{(de,ca)} \to 1/2$ (while the concurrence tends to its minimum 0, the purity tends to a asymptotic value larger than its minimum $1/4$). Thus although quantum teleportation does require the channel state to be entangled, a nonzero critical value of minimum entanglement is not always necessary [16], and this is in contrast to several previous results [2, 24–26].

As a final discussion, we would like to make some further analysis for the situation where the qubits at the transmitting station subject to external environments. By comparing the curves shown in Figs. 2a–2c, one can see that while the channel state for this case possess the largest amount of entanglement (which is exactly the same as that with only the qubit in possession of Bob subject to decoherence), it yields, however, the smallest average fidelity among the three cases considered in this Letter. Physically, one may attribute the cause of this phenomenon to the decoherence of the unknown state before it is teleported, and this is not incompatible with the fact that the purity of the channel state here is equal to or larger than those of the former two cases.

In summary, we have investigated environmental effects on fidelity of quantum teleportation by solving analytically a master equation in the Lindblad form. Different from the previous works [14–16] in which the authors considered generators of noise given by different Pauli operators, here we concentrated on the one-qubit teleportation protocol when it is executed in the presence of dissipative, noisy and dephasing environments [10]. Through detailed calculation and analysis of decay dynamics of the average fidelity with different decoherence mechanisms, we revealed that the environmental effects may cause the teleportation protocol to lose its quantum advantage over purely classical communication in a finite (e.g., the qubit of Bob or the two qubits of Alice subject to dissipative environment or the system subject to noisy environment) or infinite (e.g., both the two qubits of the channel subject to dissipative environment or the system subjects to dephasing environment) lifetime, which is dependent on the type of environment.

Moreover, we revealed that the concurrence [22] cannot completely reflect the quality of the entanglement as a resource for teleportation. The purity of the channel state is also very important. Sometimes, a channel state with less amount of entanglement can even enable teleportation with fidelity better than that with greater amount (not maximum) of entanglement (see, e.g., Fig. 2a), but now the former possess a larger amount of purity. In fact, the increase of the teleportation fidelity may be accompanied by the increase of the entanglement or the purity of the channel state or both of them. We have not found the case where the average fidelity is increased while both the entanglement and the purity of the channel state are decreased for this kind of system. For Bell states, they possess the maximum entanglement and purity and thus enables perfect teleportation. For general cases, however, how to matching these two quantities (entanglement and purity) to achieve a high teleportation fidelity is obviously vital to long distance quantum communication.

Also it is not clear whether these are the only two essential quantities for predicting fidelity or if they are specific to the channels considered. All these needs further investigation.

Before ending this Letter, we would also like to emphasize that although there are works [27, 28] showing that sometimes the teleportation fidelity may be enhanced to some extent by local environment, finding ways to minimize or even eliminate the detrimental effects of decohering environments is still the prerequisites and challenging task in the practical realization of quantum communication. Indeed, circumvention of the decoherence problem [4, 18] has been shown to be theoretically possible in various contexts. Particularly, the quantum repeater scheme [18] which combines the methods of entanglement swapping, purification and quantum memory provides an elegant solution to attack the decoherence and loss in transmission channels for long-distance quantum communication.

## Acknowledgments

This work was supported by the NSF of Shaanxi Province under Grant No. 2010JM1011, the Specialized Research Program of Education Department of Shaanxi Provincial Government under Grant No. 2010JK843, and the Youth Foundation of XUPT under Grant No. ZL2010-32.